\documentstyle[aps,epsf]{revtex}

\newcommand{\eins}{\mbox{$1 \hspace{-1.0mm}  {\bf l}$}}
\newcommand{\be}{\begin{equation}}
\newcommand{\ee}{\end{equation}}
\newcommand{\bea}{\begin{eqnarray}}
\newcommand{\eea}{\end{eqnarray}}
\newcommand{\half}{\mbox{$\textstyle \frac{1}{2}$}}

\newcommand{\shalf}{\mbox{$\textstyle \frac{1}{\sqrt{2}}$}}
\newcommand{\ket}[1]{ | \, #1  \rangle}
\newcommand{\bra}[1]{ \langle #1 \,  |}

\newcommand{\abs}[1]{ | \, #1 \,  |}

\begin{document}
\draft\onecolumn
\title{Entanglement splitting of pure bipartite quantum states}
\author{Dagmar~Bru\ss }
\address{
Institut f\"{u}r Theoretische Physik, Universit\"{a}t Hannover, Appelstr. 2,
D-30167 Hannover, Germany} \date{Received \today}
\maketitle
\begin{abstract}
The concept of entanglement splitting is introduced by
 asking whether it is possible 
for a party possessing half of a pure bipartite quantum state to  
transfer some of his entanglement with the other party to a third party.  
We describe 
 the  unitary local transformation for symmetric  
and isotropic splitting
of a singlet into two branches that leads to  the highest entanglement
of the output. 
The capacity of the resulting quantum channels is discussed.
Using the same transformation for less than maximally entangled pure states,
the entanglement of the 
 resulting states is found. We discuss whether they
can  be used to do teleportation and to test the Bell inequality. 
 Finally we generalize to entanglement
 splitting into more than two
branches. 
\end{abstract}

%\onecolumn
%\widetext
Entanglement is  a resource in 
the physics of quantum information  that has
  deserved and received increasing attention over the recent years, as
 it is a main reason for fundamental differences with classical 
 information: given a system that consists of $n$ quantum bits,
 the total state is called separable if it can
 be decomposed as
 \be
 \varrho_{sep} = \sum_i p_i \; \varrho^{(1)}_i\otimes \varrho^{(2)}_i\otimes
...         \otimes \varrho^{(n)}_i\ ,
 \ee 
 where $\varrho^{(n)}_i$ is a density matrix for the $n$th subsystem,
 and the positive weights $p_i$ obey $\sum_i p_i=1$.
 Otherwise the system is called
 entangled -- a property that exists only for quantum
 states.
 \par
Several features of entanglement have been studied so far, e.g. 
how to concentrate entanglement of several pairs into more entangled
fewer pairs
 (distillation or purification
 \cite{disti,puri}) or how to transfer entanglement
 to pairs that have not been entangled before (swapping \cite{tele}).
 It is desirable to study more properties of entanglement,
 as we know that it is the central resource for speed-ups in quantum
 algorithms, but are still  unable to fully classify multi-particle
 entanglement.
 \par
The purpose of this paper is to 
introduce the concept of entanglement splitting by
answering the question whether 
the resource of entanglement can be shared,
such that more than one party can profit from it.
Apart from the fundamental aspect of this topic
the setting has a direct relevance in teleportation \cite{tele}.
This paper  is  restricted  to
  entanglement splitting of two-dimensional
states.
\par
Let us consider the following situation: 
the two parties Alice and Bob are sharing a 
singlet\footnote{Any other Bell state would be equally good 
for our considerations.}, i.e. the Bell state
\be
\ket{\psi^-}=\shalf (\ket{01}-\ket{10})  
 \ .
\ee
 This resource can be used
%to test violation of the Bell inequality or
to do teleportation of a quantum state from one party to the other. 
The singlet can thus be seen as an ideal quantum channel along which 
quantum information can be sent.
We  ask
the following question: let us imagine 
 Bob has a brother Brian and wants to generously
share his resource with him.  There is a limit to
his generosity, though -
for example he does not want to simply swap the entanglement to Brian, 
because then
he would be left without any entanglement himself.
Can he design a unitary transformation acting
on his and his brothers state (which is supposed to be in a prescribed
state $\ket{0}$ at the beginning) and, if necessary,
 an ancilla, such that after
the transformation both Bob and Brian have 
an  identical amount of entanglement with Alice? 
Thus she could
teleport imperfectly %(otherwise she would be able to clone) 
a quantum state to both of them. 
In other words, rather than using one perfect quantum channel, there would be
two imperfect or noisy channels.
To be precise, we  would have created a  channel bifurcation
 with one input side (Alice) and two output
sides (Bob, Brian). %What is the best quality one can achieve?
% Could they then also test for multiple
%violation of Bell's inequality? 
Note that
Bob's action is local in the sense that he does
not act on Alice's side. It is not local with respect to
Brian, though.  Formally this transformation can be written as
\bea
\ket{\psi}^{in} &=& \ket{\psi^-}_{AB_1}\ket{0}_{B_2}\ket{0}_{anc}
\nonumber \\
&\rightarrow & \eins_A\otimes U_{B_1B_2anc}\ket{\psi}^{in}\ \ ,
\label{define}
\eea
where Bob is abbreviated as $B_1$ and Brian as $B_2$.
What is the highest entanglement that can remain between Alice
and Bob after his transformation?
\par
This  question has not been addressed in work about related 
areas, namely
 about broadcasting of 
entanglement \cite{knight1,knight2}, where it was shown that
an entangled pure  state can be copied
imperfectly by employing a   unitary transformation
on {\em both } subsystems and about
 telecloning \cite{vlatko}, where the authors described a multiparticle
state which allows to do imperfect
 teleportation of  a state from one party to several
others. Transformations of 
a subsystem of a maximally entangled state have 
 been used in \cite{cerf} in the context of asymmetric cloning.
 Note that in this context one could generalise our arguments to the
 case of asymmetric entanglement splitting in a straightforward way.   
\par 
In our scenario of entanglement splitting
 the emerging imperfect quantum channels can, in principle, be of any kind,
depending on Bob's transformation. 
We do not attempt a complete study here, but restrict ourselves to the
case where the ideal channel is split 
such that it can serve equally well for  teleportation of
any quantum state. The motivation for this is not to introduce any spatial
asymmetry into the splitted channel.
This means we only consider splitting of the singlet
into two identical 
branches that act as  depolarizing
channels (i.e. with the same probability of 
error in $x$-,$y$-, and $z$-direction)
when using the standard teleportation scheme.
The equivalence between quantum channels and bipartite states  has been 
studied in \cite{long,horodecki} where it was shown that a 
%depolarizing channel corresponds to a 
Werner state 
 (i.e. a state that is with probability $F_W$
in one of the Bell states and with equal probabilities $(1-F_W)/3$ in the 
remaining three Bell states)
acts as a depolarizing channel when used for standard 
teleportation.
In other words, we require the reduced density matrix of Alice and Bob 
after Bob's transformation
 to be a Werner state.
\par
The most general transformation that Bob can perform
is given by its action on the basis 
of his qubit  (Brians qubit and the ancilla are initially fixed)
and can be written as
\bea
U_{B_1B_2 anc}\, \ket{0}\ket{0}\ket{0} & = & a\ket{00}\ket{A}+b(\ket{01}
     +\ket{10})\ket{B}+ c\ket{11}\ket{C}
 \label{eq:00}\ \ ,\\
U_{B_1B_2 anc}\, \ket{1}\ket{0}\ket{0} & = & \tilde a\ket{11}
         \ket{\tilde A}+\tilde b(\ket{10}
     +\ket{01})\ket{\tilde B}
     + \tilde c\ket{00}\ket{\tilde C}\ \ .
  \label{eq:10}
\eea
where we have used the same notation as in \cite{oxibm} and
implied symmetry under exchange of Bob's and Brian's qubits.
The coefficients and ancilla states are restricted by constraints
from unitarity of $U_{B_1B_2 anc}$. We also require symmetry under
renaming the basis, namely exchange of $\ket{0}_{B_1}\leftrightarrow 
\ket{1}_{B_1}$.
Inserting (\ref{eq:00}) and (\ref{eq:10}) into equation (\ref{define}) 
and tracing over $B_2$ 
 one finds the following structure before tracing over the 
 ancilla:
 \bea
 \rho_{AB_1} &=& \frac{1}{4}\text{Tr}_{anc}\left[  \left\{
 a(\ket{\Psi^+} +\ket{\Psi^-})\ket{A}+b(\ket{\Phi^+} +\ket{\Phi^-})
     \ket{B} - \tilde b(\ket{\Psi^+} -\ket{\Psi^-})
     \ket{\tilde B}-\tilde c(\ket{\Phi^+} -\ket{\Phi^-})\ket{\tilde C}\right\}
     \left\{\bra{...}\right\} 
       \right. 
     \nonumber \\
  & & \left. \ \ + \ \ \  \left\{
 b(\ket{\Psi^+} +\ket{\Psi^-})\ket{B}+c(\ket{\Phi^+} +\ket{\Phi^-})
     \ket{C} - \tilde a(\ket{\Psi^+} -\ket{\Psi^-})
     \ket{\tilde A}-\tilde b(\ket{\Phi^+} -\ket{\Phi^-})\ket{\tilde B}\right\}
     \left\{\bra{...}\right\}  
 \right]\ \ .
 \label{tedious}
 \eea 
 Here the notation $\left\{\bra{...}\right\}$ indicates that the whole ket-vector 
 to the left
 now appears as bra-vector.
 We have used the customary definition for the Bell states, namely
 \bea
 \ket{\Psi^\pm} &=&\shalf (\ket{01}\pm\ket{10}) \ \ ,\nonumber \\
 \ket{\Phi^\pm} &=&\shalf (\ket{00}\pm\ket{11})\ \ .
 \label{bell}
 \eea
 After tracing over the ancilla we have to set (\ref{tedious})
  equal to the Werner state we aim at,
 namely 
 \be
 \rho_{AB_1}= F_W \ket{\Psi^-}\bra{\Psi^-}+\frac{1-F_W}{3}
     (\ket{\Psi^+}\bra{\Psi^+}+\ket{\Phi^+}\bra{\Phi^+}+
     \ket{\Phi^-}\bra{\Phi^-})\ \ .
  \label{wstate}
 \ee
 The explicit calculation is laborious and therefore not shown
 explicitly.
 It yields some constraints
 for the parameters 
characterizing
this transformation:
\begin{itemize}
\item[(i)]$\abs{a}^2-\abs{c}^2=\abs{\tilde a}^2-\abs{\tilde c}^2
    %=\third (4 F_W-1)
    $
\item[(ii)]$\abs{a}^2-\abs{c}^2=\mbox{Re}
        \left[ \tilde b^* a \bra{\tilde B}A\rangle
            +\tilde a^* b \bra{\tilde A}B\rangle \right] $
\item[(iii)]Im$\left[   \tilde b^* a\bra{ \tilde B}A\rangle
            +\tilde a^*  b \bra{\tilde A} B\rangle \right] = 0$
\item[(iv)]$b^* \tilde c \bra{ B}\tilde C\rangle
            +c^* \tilde b \bra{C}\tilde B\rangle = 0$
\item[(v)]$b^* a\bra{ B}A\rangle
            + c^* b\bra{C} B\rangle= 0 $
\item[(vi)]$ \tilde b^* \tilde a\bra{ \tilde B}\tilde A\rangle
            + \tilde c^* \tilde b\bra{\tilde C} \tilde B\rangle=0$
\item[(vii)]$ \tilde c^*a \bra{ \tilde C}A\rangle
            -\tilde a c^*\bra{C} \tilde A\rangle = 
            \tilde b b^* \bra{ B}\tilde B\rangle-b \tilde b^* 
            \bra{ \tilde B}B \rangle$
\end{itemize}
It turns out that these constraints coincide 
-- apart from (vii) --
with 
the ones the parameters of an isotropic and symmetric 
$1\to 2$ quantum cloning transformation have to fulfil. 
(These are given in 
section IIA of \cite{oxibm}.)
This  is not astonishing, as
in both scenarios we want to find a transformation with the 
highest possible degree of symmetry. Nevertheless is not obvious, either,
as in this paper we perform a transformation on a part of
an entangled state, i.e. a system in which the reduced density matrix 
 is the identity and does not contain  
information   about any direction of polarization,
in contrast with the scenario in quantum cloning. As mentioned above, the 
sets of constraints in the two scenarios are not exactly identical.
\par An isotropic $1\to 2$ cloner was introduced by 
Bu\v{z}ek and Hillery \cite{buzek}
and  shown to be optimal in \cite{oxibm,gisin}.
A less than optimal but still isotropic cloning 
transformation can be constructed by varying not only the coefficients
of the unitary transformation,
but also the scalar products of the ancilla, such that the conditions
given in \cite{oxibm} are preserved. 
(Note that for the  case of less than optimal cloning
the ancilla dimension has
to be increased to at least two qubits.) This statement is best 
illustrated by providing an example. 
The  fidelity $F_C = \bra{\psi^{in}}\rho^{out}\ket{\psi^{in}}$ 
of a cloning transformation 
is a measure for how close
the reduced density matrix $\rho^{out}$
of a clone is to the input. With the 
following bad cloner we reach only $F_C=3/4$:
\bea
U^{\text{bad}}\ket{0}\ket{0}\ket{0}_{a1} \ket{0}_{a2}
     &=&\shalf \ket{00}\ket{0}_{a1}\ket{0}_{a2}+
    \half (\ket{01}+\ket{10})\ket{1}_{a1}\ket{0}_{a2}   \nonumber \\
 U^{\text{bad}}\ket{1}\ket{0}\ket{0}_{a1}\ket{0}_{a2}
     &=&\shalf \ket{11}\ket{1}_{a1}
      \shalf(\ket{0}_{a2}+\ket{1}_{a2})+
    \half (\ket{01}+\ket{10})\ket{0}_{a1}\shalf(\ket{0}_{a2}+\ket{1}_{a2})  \ \ ,
\eea
where the two first bits on the right hand side are the clones and
the subscripts $a1, a2$ denote the ancilla states.
\par
Which of the many possible isotropic cloning machines optimizes
 the quality of the channels resulting from 
entanglement splitting? 
From equation (\ref{tedious}) and the 
 constraints given above one finds that  
 the fraction of the singlet in our output channel is
 \be
 F_W=\frac{1}{4}[3(\abs{a}^2-\abs{c}^2)+1] \ \ ,
 \ee
 which is
related to the cloning fidelity $F_C$ in a simple way, namely
\be
F_W=\half (3F_C-1) \ .
\ee 
Remember that our constraint (vii) is slightly different from the
according constraint in cloning. When going through the maximisation procedure,
given in the appendix of \cite{oxibm}, step by step, it turns out,
however, that the
%In addition, 
%the same transformation that minimizes the cloning fidelity
%also minimizes $F_W$. 
 same transformation that maximizes the cloning fidelity
also maximizes $F_W$ in our scenario.
The upper bound of a universal cloner is given by
$ F_C \leq 5/6$, 
%(the upper bound is given in \cite{oxibm} and the lower one in \cite{gipo})
and therefore  the Werner
fidelity has to fulfil $F_W \leq 3/4$.
The entanglement of a Werner state with $F_W\geq 1/2$
is an increasing function of $F_W$.
In other words, the highest entanglement
of the output is reached when Bob performs an
optimal  cloning transformation on his and Brian's qubit.
\par 
 It is interesting to observe that our method provides
    a simple derivation for the upper limit of the
quality of the universal NOT or spin-flip
(which is equal to the lower bound on 
the cloning fidelity): as $F_W=\bra{\Psi^-}\rho_{AB}\ket{\Psi^-}$ 
has to be positive, we arrive at $F_C\geq 1/3$ which corresponds to the
 upper bound for the universal NOT given in \cite{gipo}. In other words, 
 when applying our  positive map it is enough  to require 
 positivity of $\rho_{AB}$ in order to find an explicit bound for the
 best universal NOT
 (when the corresponding  transformation acts on  Bob's qubit it
erases the singlet completely).
 \par
 We have shown that the best way to split  entanglement is to use
 the optimal  universal cloner. 
Note that even a global unitary transformation acting on Alice's, Bob's and
Brian's bit cannot split the singlet into 
depolarizing channels with higher degree
of entanglement  because 
Alice could then teleport a state to both 
Bob and Brian with higher quality than the optimal 
 cloning quality.
 It was shown in \cite{oxibm} that the quantum capacity
 of  the depolarizing channels that 
 can be reached in this scenario  vanishes. 
In other words: it is impossible to bifurcate
 a perfect quantum channel into two depolarizing channels
with non-vanishing capacity.
  \par
Note that this observation sets
limits on the possible amount of entanglement of subsystems in multipartite
states with the described structure, see also the 
generalisation to splitting into more than two branches given below. 
\par
Let us now look at the case where Alice and Bob share a pure, but not 
maximally entangled state of the kind
\be
\ket{\psi^-(\alpha)}=\alpha \ket{01}-\beta\ket{10} \ \ \ \text{with} 
\ \ \ \abs{\alpha}^2+\abs{\beta}^2=1\ .
\label{alpha}
\ee
Note that {\em every} pure bipartite state can be written in
this way by using the Schmidt decomposition
and naming the bases of the two systems accordingly.
If Bob performs the same optimal cloning transformation as above 
(note that the
original channel is not an ideal channel and the outcomes are no 
depolarizing channels, so our optimality arguments considered above
do not apply),
  the reduced density matrix of Alice and Bob
(or Alice and Brian) after this operation is
\be
\rho_{AB} = \frac{1}{6}
\left( \begin{array}{cccc} \abs{\alpha}^2& 0 & 0 & 0 \\
                             0 & 5\abs{\alpha}^2 & -4\alpha\beta^* & 0 \\
                             0 &  -4\alpha^*\beta&  5\abs{\beta}^2 &0 \\
                             0 & 0 & 0 &  \abs{\beta}^2\end{array}\right)
\label{density}
\ee
in the basis $\{\ket{00},\ket{01},\ket{10},\ket{11}\}$.
Is this matrix inseparable?
Using the separability criterion established by Peres \cite{peres} and
the Horodeckis \cite{horo} one has to check the partially
transposed density matrix for positivity. Its eigenvalues
are
\bea
\lambda_1 & = & \frac{5}{6}\abs{\alpha}^2 \ \ ,\nonumber \\
\lambda_2 & = & \frac{5}{6}\abs{\beta}^2 \ \ ,\nonumber \\
\lambda_{3,4} & = & \frac{1}{12}(1\pm \sqrt{1+60\abs{\alpha}^2 
                           \abs{\beta}^2}) \ \ ,
\eea
and as $\lambda_4<0$ for any $\alpha$ with 
 $\abs{\alpha}\neq 0,1$, the output is inseparable for {\em any}
entangled input. So, no matter how small Bob's entanglement
was at the beginning, he can still make a donation of it
to Brian. Something like a smallest unit of entanglement, 
that could not be split, does not exist.
\par
How much entanglement remains in the state of Alice and Bob
after the transformation?
The entanglement of formation for a given density matrix $\rho$
was introduced in \cite{wootters}
and is given by
\be
E(\rho)=-(\half +\half \sqrt{1-C^2})\log_2(\half +\half \sqrt{1-C^2})-
          (\half -\half \sqrt{1-C^2})\log_2(\half -\half \sqrt{1-C^2})
\ee
 where  the concurrence $C$   
 is defined as
 \be
C(\rho)= \text{max}\{0,\sqrt{\xi_1}- \sqrt{\xi_2}- 
     \sqrt{\xi_3}-\sqrt{\xi_4} \}  \ \ 
\ee
with $\xi_i$ being the eigenvalues of
$\rho[(\sigma_y\otimes\sigma_y) \rho^*(\sigma_y\otimes\sigma_y)]$,
ordered by size as $\xi_1\geq \xi_2\geq 
     \xi_3\geq\xi_4$.
 \par
\begin{figure}[hbt]
\setlength{\unitlength}{1pt}
\begin{picture}(500,300)
\put(360,-170){$\abs{\alpha}$}
\put(0,140){EOF}
\epsfysize=10cm
\epsffile[72 230 540 560]{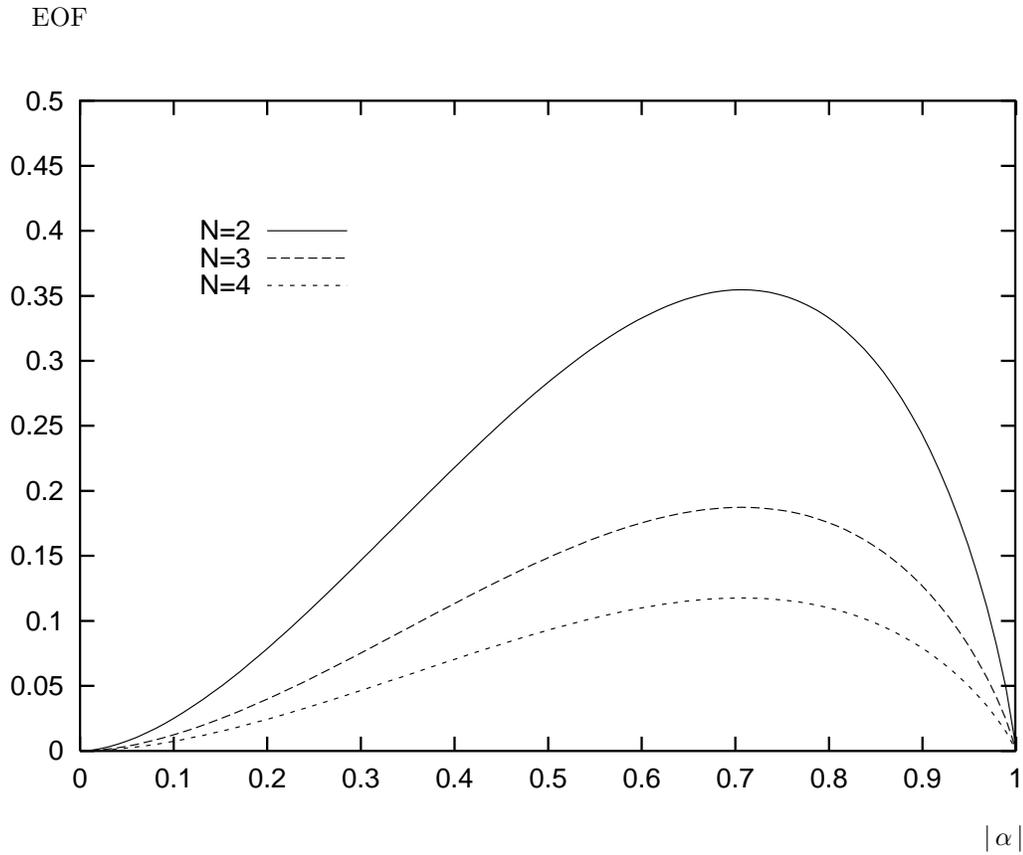}
\vspace{-0.2cm}
\end{picture}
\vspace{7cm} 
\caption[]   
        {\small Entanglement of formation of any of the resulting branches
        %after entanglement splitting 
        as a function of initial parameter $\abs{\alpha}$, see equation
        (\ref{alpha}), for different numbers of branches $N$.
               }
\label{figure1}
\end{figure}

The entanglement of formation for the density matrix in (\ref{density}),
i.e. the density matrix of Alice and Bob
or Alice and Brian after the transformation is
found by calculating the concurrence as
 \be
C(\rho)=\abs{\alpha\beta}       \ \ .
\ee
Note that the concurrence of the original state was
$C_{orig}=2\abs{\alpha\beta}$.
We observe that $C(\rho)>0$ for $0<\abs{\alpha}<1$. A
discussion about inequalities for the squares of the concurrences
in the subsystems of three-particle states is presented in
\cite{wootters2}.
 The curve for the entanglement of formation is shown in figure \ref{figure1}.
 \par
\begin{figure}[hbt]
\setlength{\unitlength}{1pt}
\begin{picture}(500,300)
%\put(360,-170){$\alpha$}
%\put(0,140){EOF}
\epsfysize=10cm
\epsffile[72 230 540 560]{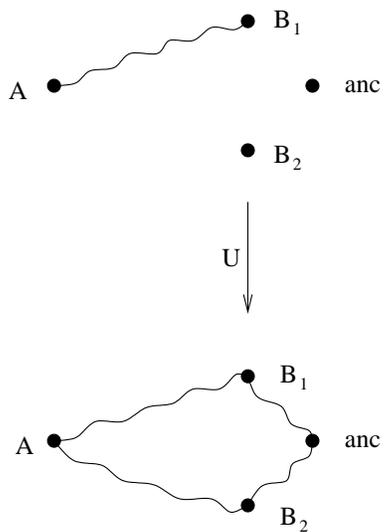}
\vspace{-0.2cm}
\end{picture}
\vspace{3cm} 
\caption[]   
        {\small Entanglement of the total state before and after
        splitting a pure entangled bipartite state into two branches.
         The wavy lines between two parties
        indicate that their reduced density matrix is entangled.  
        %Parties that are not connected by a wavy line share a separable
        %density matrix.
               }
\label{entangle}
\end{figure}

 There is some loss of entanglement 
  in the system of Alice, Bob and Brian 
  compared to the original entanglement because the 
  unavoidable ancilla is entangled  as well.
 Which subsystems of the total state after the 
 transformation will be entangled with each other? By looking at the
 density matrices for the bipartite subsystems one finds the following:
 $A$ is entangled with $B_1$ and with $B_2$. The ancilla is entangled
 with $B_1$ and with $B_2$. $B_1$ and  $B_2$ are not entangled with 
 each other. This can be visualized as in figure \ref{entangle}.
\par
Is the remaining entanglement between
Alice and each of the brothers enough 
so that she can do teleportation to both for {\em any} initial $\alpha$?
We learned in \cite{horo2} that a mixed state is better for teleportation
than  classically
whenever the following criterion for the maximally
reachable fidelity
is fulfilled:
\be
F_{max}=\half (1+\frac{1}{3} \text{Tr} \sqrt{T^\dagger T})>\frac{2}{3}
\ee
where 
the matrix $T$ is defined  by  the  general expansion of 
the density matrix
\be
\rho = \frac{1}{4}(\eins \otimes \eins +\vec s^{(1)}\cdot\vec \sigma \otimes
\eins       +\eins\otimes\vec s^{(2)} \cdot\vec \sigma + T_{ij} \sigma_i
\otimes\sigma_j) \ \  
\ee
and in our case given by 
\be
T = \frac{2}{3}
\left( \begin{array}{ccc} -2\text{Re}[\alpha\beta^*]& -2\text{Im}[\alpha\beta^*]& 0  \\
                        -2\text{Im}[\alpha\beta^*] & -2\text{Re}[\alpha\beta^*]  & 0 \\
                             0 & 0  &  -1\end{array}\right)\ \ 
\label{tmatrix}
\ee
in the basis $\{x,y,z\}$.
\par
In order for the teleportation scheme to work we thus find the following 
window for the initial state:
\be
\frac{1}{2}(1-\frac{\sqrt{15}}{4})<\abs{\alpha}^2<
\frac{1}{2}(1+\frac{\sqrt{15}}{4})
\ \ .
\ee
This is a very wide range for $\alpha$, namely
$0.008< \alpha < 0.992$. Thus, even for small initial entanglement
the teleportation fidelity of the splitted channel is still higher than 
the classical fidelity. 
\par 
Could we also use the output state to do a three-party test of the 
Bell-CHSH
inequality \cite{chsh}?
 Using the criterion given in \cite{horo3,horo4}, we have to
calculate the eigenvalues of $T^\dagger T$ where $T$ is given in equation
(\ref{tmatrix}). The authors showed that the Bell inequality
is equivalent to the sum of the two largest eigenvalues being smaller
or equal to one. In our example this is the case for any $\alpha$, 
so our outputs do not violate the Bell inequality. The  output 
state is another example (for $\abs{\alpha}\neq\abs{\beta}$
 distinct from the ones given in 
\cite{horo4}) for the fact that separability is a different criterion
from violation of Bell's inequalities.
\par 
In the remaining part of the paper we want to generalize our results to 
the case where Bob is willing to share part of his entangled state,
given in equation (\ref{alpha}),  with 
 more than one brother. Let us denote the number of brothers by N-1. Thus 
N refers to the number of output channels including Bob.
Making use of the  $1 \to N$ cloning transformation given in 
\cite{gisin} we can  generalize to
 entanglement splitting into N branches. 
In this case the ancilla swallows a higher amount of entanglement.
Therefore the interesting questions here
are: is there a threshold number $N$ where not every entangled input 
leads to entangled outputs? Is teleportation 
through the output channels still possible for any number $N$,
and what is the condition  the initial entanglement has to fulfill?
  \par
Explicitly we find the following results. 
The reduced density matrix of Alice and Bob (or any of his brothers)
is given by
\be
\rho_{AB}(N) = \frac{1}{3N}
\left( \begin{array}{cccc} 
             (N-1)\abs{\alpha}^2& 0 & 0 & 0 \\
              0 & (2N+1)\abs{\alpha}^2 & -(N+2)\alpha\beta^* & 0 \\
              0 &  -(N+2)\alpha^*\beta&  (2N+1)\abs{\beta}^2 &0 \\
              0 & 0 & 0 &  (N-1)\abs{\beta}^2\end{array}\right)\ \ .
\label{densityN}
\ee
The concurrence 
of this density matrix is 
\be
C(\rho)=\frac{2}{N}\abs{\alpha\beta}      \ \ ,
\ee
i.e. %it decreases linearly with $N$, and 
the output concurrences of 
Alice with all
brothers sum up to the input concurrence
between Alice and Bob.
The according entanglement of formation of one of the $N$ branches is 
shown in figure \ref{figure1} for $N=3$ and $N=4$: we find that
 the output states are inseparable for {\em any} $\alpha$ with
 $\abs{\alpha}\neq 0,1$ and for 
{\em any} $N$;  Bob can give as many brothers as he wants 
some of his entanglement, no matter how much he had to start with.
For $N\to \infty$ the output entanglement goes to
zero for any $\alpha$.
\par
The condition the initial state has to fulfill in order to make 
teleportation through the output channels possible is
\be
\frac{1}{2}(1-\frac{\sqrt{3(2N+1)}}{(N+2)})<\abs{\alpha}^2<
\frac{1}{2}(1+\frac{\sqrt{3(2N+1)}}{(N+2)})
\ \ .
\ee
The window for teleportation shrinks like
$ 1/\sqrt{N}$ to zero width for $N\to \infty$.
\par
We found that entanglement splitting is an ``easy'' task, in the sense
that we only need to apply the optimal cloning transformation,
which splits any initial pure state entanglement. Does the
same statement also hold for an initial mixed state? It does not, as can
be seen  by giving a counter example. Let us consider the same scenario
as above, but now Alice and Bob share an initial Werner state
with fidelity $F_W$ and 
of the form given in equation (\ref{wstate}). If Bob simply applies the 
cloning transformation as for the case of a pure input, this leads to an 
output density matrix for Alice and Bob
that is separable for $1/2 \leq F_W \leq 5/8$,
a parameter region in which the input was entangled. 
In this sense mixed state entanglement
is qualitatively different from pure state entanglement. It is still an open
question, though, whether there exists a different transformation 
that could split any initial mixed state entanglement.
\par
To summarize, 
we have introduced the concept of entanglement splitting.
We have shown that the well-known 
optimal cloning transformation 
also maximizes the entanglement after 
 symmetric  splitting
of a singlet into two branches. 
The
capacity of the resulting depolarizing channels vanishes.
%The singlet cannot be split symmetrically such that  both
%branches violate Bell's inequality.
We
observed that {\em any} pure entangled state of two qubits 
can be split such
that after the transformation there is some remaining entanglement.
Teleportation, though,  can   only be performed 
better than classically if
 the entanglement of the original state exceeds a certain threshold. 
We have  generalized the results to the case of
splitting into more than one branch. In this 
scenario we still find non-vanishing
resulting entanglement for any initial entanglement and 
{\em any} number of branches $N$, tending to zero for 
infinitely many branches. Teleportation is possible for any 
$N$, if the initial entanglement is higher than a
 threshold that depends on $N$.
We hope that the ideas developped in this paper 
help to understand some fundamental aspects of
entanglement. 
%and will prove to be
%useful for other  applications in multiparty tasks.
   \par Helpful
discussions with Vlatko Vedral 
 are gratefully
 acknowledged. This work was partly completed with
 support  by the European TMR Research
Network ERP-4061PL95-1412, partly with support from Deutsche
Forschungsgemeinschaft under grant SFB 407.

\newpage     
\vspace{-2cm}

\end{document}